\documentstyle[epsf,aasms]{article}
\tighten
\begin{document}

\title{A new class of unstable modes of rotating relativistic stars}

\lefthead{Andersson}
\righthead{A new class of unstable modes}

\author{Nils Andersson}

\affil{Department of Physics,  Washington University, 
St Louis MO 63130, USA}

\abstract
The first numerical study of 
axial (toroidal) pulsation modes of a slowly
rotating relativistic star is presented. The calculation includes terms of 
first order in $\epsilon \equiv \Omega
\sqrt{R^3/M}<<1$ ($R$ is the radius, $M$ is the mass
and
$\Omega$ is the rotation frequency of the star), and accounts for
effects due to the coriolis force. Effects due to the centrifugal 
flattening of the star enter at order $\epsilon^2$ and are 
 not included in the analysis. It is shown that increased
rotation tends to decrease the damping times for prograde modes,
 while retrograde become longer lived. Specifically, we show that 
rotation affects the axial gravitational-wave $w$-modes 
in this way. We also present the first
relativistic calculation of the so-called $r$-modes (analogous to 
Rossby waves in the Earth's oceans). 
These have frequencies of the same order of magnitude as the rotation frequency
of the star. The presented results indicate that the $r$-modes are unstable due to 
the emission of gravitational radiation for \underline{all} rotating perfect
fluid stars.
This is interesting since the previously considered gravitational-wave
instability associated with (for example) the $f$-mode of the star sets 
in at
a critical rotation rate. Because they are unstable also for the slowest 
rotating stars the $r$-modes may well be of
considerable astrophysical importance.
\endabstract


\section{Introduction}

This paper is the first in a series of
investigations of perturbations of slowly rotating relativistic stars. 
The original intention was to present an exhaustive 
description of the problem, together with a detailed discussion of
the effect that rotation has on the various pulsation modes of
a relativistic star. However, a preliminary investigation
into the axial (also referred to as toroidal, or odd-parity)
mode-problem unveiled an interesting and rather 
surprising result: There exist pulsation modes of
a rotating relativistic star that are unstable due to the emission 
of gravitational
radiation at \underline{all} rates of rotation. That is, 
these modes --- the relativistic analogue
of the Newtonian $r$-modes, cf. Papaloizou and Pringle (1978)
and Saio (1982) --- are 
unstable even for very slowly rotating perfect fluid stars. 
The instability of the $r$-modes is thus different to previously
considered mode-instabilities  (e.g. for the
$f$-mode of the star) that set in at a certain rate of rotation (Comins 1979; Friedman 1983; Managan 1985; Imamura, Durisen and Friedman 1985;
Lindblom 1986; Lindblom and Mendell 1995; 
 Stergioulas and Friedman 1997). 
But the fact that the $r$-modes are unstable can still 
be understood in terms of 
the mechanism that was first discussed
by  Chandrasekhar (1970) and Friedman and Schutz (1978). 

This paper describes the calculation 
that led to the discovery of the unstable $r$-modes 
without many of the technical details. These will be reported
elsewhere.

\section{Slowly rotating relativistic stars}

This paper concerns pulsation modes of a
slowly rotating relativistic star. The modes follows from a
study of linear perturbations of the stellar fluid and the
associated spacetime metric, for a star that
rotates sufficiently slowly that the effect of rotation can be 
considered as a small perturbation of a non-rotating configuration. 
The analysis is based on 
a double perturbation expansion. To lowest 
order the star is static and non-rotating, and the
perturbations are such that
$|h_{\mu \nu}|<<1$ , where $g_{\mu \nu} = g_{\mu \nu}^{\rm background} +
h_{\mu\nu}$, and also $\epsilon \equiv \Omega
\sqrt{R^3/M}<<1$. Here $R$ is the radius of the star, $M$ is the mass,
and
$\Omega$ is the rotation frequency of the star according to 
a distant observer (the star rotates uniformly so $\Omega$ is a constant).
Geometrized
units $c=G=1$ are used throughout the paper.

Before presenting the details of the problem it is meaningful
to discuss whether a  ``slow-rotation expansion'' is likely 
to provide results of physical relevance. It is rather
straightforward to show that this is the case.
The rotation of a star is absolutely limited by the Kepler frequency, at which
mass shedding at the stellar equator makes the star unstable. 
The following empirical formula has proved
to be a reasonable approximation of this limiting frequency 
\begin{displaymath}
\Omega_K \approx C \sqrt{ {M \over M_\odot} ({10 \mbox{ km} \over
R})^3 } \ ,
\end{displaymath}
where $C \approx 7.8\times 10^3 \mbox{s}^{-1}$ (Haensel and Zdunik 1989;
Friedman, Ipser and Parker 1989; Cook, Shapiro and Teukolsky 1994).
$M$ and $R$ refer to the 
mass and the radius
of the corresponding non-rotating  star. 
By rewriting the definition
of the expansion parameter $\epsilon$ in the spirit of this relation
we get
\begin{displaymath}
\epsilon =\Omega D \sqrt{ {M_\odot \over M} ({R\over 10
\mbox{ km}})^3 }  \ ,
\end{displaymath}
where $D = 8.6\times10^{-5} $ s.  That
is, the limiting Kepler frequency corresponds to a value of $\epsilon$ that
is slightly smaller than $0.7$. This is the largest 
value of $\epsilon$ that should ever be relevant. A calculation 
including only terms of order $\epsilon$ will probably not provide an accurate
description of extremely rapidly rotating stars, but it is interesting 
to note that an approach that is consistent to $O(\epsilon^2)$ may 
prove useful also for such cases. It has, in fact, been shown that 
a second-order calculation ($\epsilon^2$) 
of the upper mass limit (and other properties) for rotating neutron
stars is accurate to within a few percent (Weber and Glendenning 1992). 

As a second example, consider a millisecond pulsar with 
a period of 2 ms (cf. PSR 1937+21 that has a 
period of 1.6 ms). This corresponds to 
$\Omega = 3100 \mbox{s}^{-1}$ and if we for simplicity assume that 
$R=10 \mbox{ km}$ and $M=M_\odot$ 
the corresponding value of $\epsilon$ is roughly $0.3$. 
This value should be within reach of a calculation
to first order in the rotation expansion. Consequently, 
such a calculation has the potential to provide considerable
information about the fastest spinning pulsars that have been 
observed.

There are practical reasons why the perturbation approach may be preferred
to a fully nonlinear calculation. There are by now several reliable 
methods for numerically solving the Einstein equations to
 construct fully relativistic rotating stellar models
(see, for example, Friedman et al. 1986; 
Stergioulas and Friedman 1995), but to calculate the
neutral modes (that signal the onset of gravitational-wave instability) for
these models is still
a difficult task (Stergioulas and Friedman 1997). And it will be considerably harder to calculate the actual pulsation modes of a rapidly rotating, 
relativistic star. 
A full consideration of the pulsation properties of rapidly 
 rotating stars will require new computational techniques and considerable
 computer power. In contrast, the slow-rotation approach should not 
require any conceptually new steps. The relevant perturbation equations can
be reduced to wave equations, and the main difference from the non-rotating case is
that one must consider several coupled equations. That the equations
take a standard form is an enormous advantage when one is interested
in extracting the complex frequencies of the pulsation modes. 
Moreover,  the slow-rotation 
calculation serves as a useful benchmark test for future  work on rapidly
rotating stellar models.  

\section{Axial perturbations of a slowly rotating star}

As was shown by, for example, Hartle (1967), the centrifugal
force affects the shape of a rotating relativistic
star through terms of order $\epsilon^2$. Hence, one can 
assume that a slowly rotating star remains spherical as a first approximation. 
The corresponding metric can be written
\begin{equation}
ds^2 = - e^\nu dt^2 + e^\lambda dr^2 + r^2(d\theta^2 -2\omega \sin^2 \theta dt
d\varphi + \sin^2 \theta d\varphi^2)
\label{metric}\end{equation}
For a given equation of state $p = p(\rho)$, where $\rho$ is 
the energy density, the pressure $p$ and the
two metric coefficients $\nu$ 
and $\lambda$ are determined by the standard TOV equations [I use 
standard notation and the relevant equations are listed by, for example,
Kojima (1992)].  

When terms of order $\epsilon$ are included in the slow-rotation
expansion one must account for the  ``dragging of inertial frames'',
that is represented by $\omega$ in (\ref{metric}). After 
defining
$\tilde{\omega} = \Omega - \omega$ we have (Hartle 1967)
\begin{equation}
{ d^2 \tilde{\omega} \over dr^2} - \left[ 4\pi(\rho+p)e^\lambda r - 
{4\over r} \right] { d\tilde{\omega} \over dr} - 16\pi (\rho+p) e^\lambda 
\tilde{\omega} = 0 \  .
\end{equation}
The desired solution to this equation is well behaved at the centre
of the star; 
\begin{equation}
\tilde{\omega} \approx \tilde{\omega}_0 \left[ 1 + {8\pi \over 5} (p_0 + \rho_0) r^2 +
O(r^4) \right] \ , \quad \mbox{ as } r \to 0 \ ,
\end{equation}
where $p_0$ and $\rho_0$ are the central values of the pressure and the density. 
In the vacuum outside the star we have
\begin{equation}
\omega = {2J \over r^3}
\end{equation}
where $J$ is the total angular momentum of the star. The constants  
$\tilde{\omega}_0 $ and $J$ are fixed by the requirement that $\tilde{\omega}$, 
and its derivative, be continuous across
$r=R$. 

The equations that describe perturbations of a slowly rotating star
in general relativity have previously been considered by Chandrasekhar
and Ferrari (1991a) and Kojima (1992). I will
use the notation of Kojima (1992), and refer the interested reader
to his paper for a listing of the 
original perturbation equations. 

The perturbations of a non-rotating star can be divided into
two classes: axial and polar (often referred to as odd- and even
parity perturbations, respectively). The equations that describe
these two classes decouple and can
 be considered separate from each other.
Specifically, one can show that axial perturbations 
are governed
by a single wave equation (in Regge-Wheeler gauge)
\begin{equation}
{\cal D}_l^X(X_{lm}) = 
{d^2 X_{lm} \over dr_\ast^2} + (\sigma^2 - V^X_l )
X_{lm} = 0 \ .
\label{wavq}\end{equation}
Once the function $X_{lm}$ is known one can infer the
metric perturbations $h_{t\theta}$, $h_{t\varphi}$, 
$h_{r\theta}$ and $h_{r\varphi}$ as well as the fluid displacement 
$\xi^\varphi$. The angular dependence of all
perturbed variables is expressed in terms of the spherical 
harmonics $Y_{lm}$ [for details, see Kojima (1992)].
In (\ref{wavq}) it is assumed that the time dependence of the perturbation
is $e^{-i\sigma t}$. The assumption of a harmonic time-dependence is adequate 
as long as the focus is on the spectral properties of the star.
The tortoise coordinate $r_\ast$, 
which is defined by
\begin{equation}
 {d\over d r_\ast} = 
 e^{(\nu -\lambda)/2}{d\over dr} \ ,
\end{equation}
is familiar from studies of perturbed black holes, 
and the effective potential is
\begin{equation}
V^X_l (r) = e^\nu \left[ {l(l+1) \over r^2 } - { 6M\over r^3 } + 4\pi(\rho-p)
\right] \ . 
\end{equation}
This expression is valid both inside the star and in the exterior
vacuum [$M(r)$ is the mass inside radius $r$].

Let us now  assume that the star is slowly rotating.  To 
incorporate the first order effects of rotation on the
perturbations we consider
\begin{equation}
X_{lm} = X^{(0)}_{lm} + X^{(1)}_{lm} + ...
\end{equation}
where $X^{(0)}_{lm}$ is a solution to the axial equation for a nonrotating star
and $X^{(1)}_{lm}$ corresponds to terms of order $\epsilon$.
This results in the coupled equations 
\begin{equation}
{\cal D}_l^X(X^{(0)}_{lm}) = 0 \ ,
\label{ax0}\end{equation}
and
\begin{equation}
{\cal D}_l^X(X^{(1)}_{lm}) = {\cal A}^{X}_{lm} 
+  {\cal P}^{X}_{l+1m} + 
{\cal P}^{X}_{l-1m}	\ . 
\label{ax1}\end{equation}
In this schematic description, ${\cal A}^{X}_{lm} $ represents the
coupling to the zero order axial equation, and ${\cal
P}^{X}_{l\pm1m} $ couple the first order axial perturbations to the 
zeroth order polar ones.

The explicit form of the coupling to the lowest order axial solution is
(Kojima 1992)
\begin{eqnarray}
{\cal A}^{X}_{lm} &=& {2m \over l(l+1)\sigma} 
\left\{ \left[  l(l+1)\sigma^2 (\Omega - \tilde{\omega})   
+ [l(l+1)-2]8\pi(p+\rho)e^\nu\tilde{\omega} \right. \right. \nonumber \\
&-& 
\left.  { e^\nu \over r^2} {d\tilde{\omega} \over dr} 
\left\{ [l(l+1)-2](2r-5M-4\pi pr^3) - 2re^{-\lambda} \right\} \right] 
X^{(0)}_{lm} \nonumber  \\
&+& \left. l(l+1) e^{(\nu-\lambda)/2}{d\tilde{\omega} \over dr}  
{d X^{(0)}_{lm} \over dr_\ast } \right\} \ .
\end{eqnarray}
In the exterior vacuum this expression reduces to
\begin{eqnarray}
{\cal A}^{X}_{lm} &=& {2mJ \over l(l+1)\sigma} 
\left\{ \left[ { 2l(l+1)\sigma^2 \over r^3 } 
- 6 e^\nu { [l(l+1)-3](2r-5M) -M \over r^6} \right] X^{(0)}_{lm}
\right. \nonumber \\
&+& \left. {6l(l+1)e^\nu\over r^4}  {d X^{(0)}_{lm} \over
d r_\ast } \right\} \ .
\end{eqnarray}
The explicit form for ${\cal P}^{X}_{l\pm1m} $ was given by Kojima (1992).

The apparent coupling
between the axial and polar perturbations in (\ref{ax1}) potentially 
complicates the 
definition of a stellar pulsation mode, but these
complications enter only at order $\epsilon^2$ and higher.  
As in the non-rotating case, the
asymptotic behaviour of a general solution is
\begin{equation}
X_{lm} \sim x_{\rm out} e^{i\sigma r_\ast} + x_{\rm in} e^{-i\sigma r_\ast} 
\end{equation}
for a general frequency $\sigma$. 
Here $x_{\rm out} = x_{\rm out}^{(0)} + x_{\rm out}^{(1)}$, and similar
for $x_{\rm in}$ . 
In analogy with the non-rotating case 
a quasinormal mode of the star
should correspond to a frequency $\sigma_n$ such that 
$x_{\rm in}(\sigma_n) = 0$.
In principle, one can find such solutions to the axial slow-rotation 
equations for any magnitude of ${\cal P}^{X}_{l\pm1m} $ 
(Chandrasekhar and Ferrari 1991a).
But in this paper the primary interest concerns modes that limit to the axial
modes of a nonrotating star as $\epsilon\to 0$. For such modes the 
zero order (nonrotating) polar perturbations should vanish, and hence one should
use ${\cal P}^{X}_{l+1m} = {\cal P}^{X}_{l-1m} = 0$ in (\ref{ax1}). The
coupling between the two classes of perturbations will 
affect the mode-frequencies only
at order $\epsilon^2$ and higher (Kojima 1993a). 
In principle, an axial mode  generates 
a polar perturbation
of order $\epsilon$ that feeds back and affects the axial 
mode-spectrum at second order.
Since this paper is restricted to the 
first order axial mode-spectrum the effect of the 
coupling to the polar perturbations will not be included. 
An axial pulsation mode is
defined by a solution to (\ref{ax0}) and (\ref{ax1}), with 
${\cal P}^{X}_{l+1m} = {\cal P}^{X}_{l-1m} = 0$, for which 	
$x_{\rm in}(\sigma_n) = 0$. 
This approach is identical to that used by Kojima for the polar 
problem (Kojima 1993b). The idea seems
consistent, but the definition
of modes for a rotating star is a complicated issue that 
requires further discussion.

The integration of (\ref{ax0}) and (\ref{ax1}) is initiated
at a small value of $r$ using the regular power series solutions. In the 
case of (\ref{ax1}) I use only the particular solution that 
corresponds to the coupling term ${\cal A}^{X}_{lm}$.  
The equations are then integrated to
the surface of the star $r=R$. This part of the numerical calculation is
straightforward. The main difficulty arises in the exterior of the star,
and is associated with the well-known fact that the eigenfunction of
a pulsation mode  diverges as $r\to \infty$. Specifically, a 
purely outgoing-wave solution will behave as $X_{lm}\sim e^{i\sigma r_\ast}$ 
for large $r$. A pulsation mode that is damped with time according to an 
observer at fixed $r$ corresponds to
$\mbox{Im } \sigma_n < 0$, and the desired
solution to (\ref{ax0}) and (\ref{ax1})
will therefore diverge as $r$ increases. For a general complex frequency the 
decaying  solution  --- that represents ingoing waves --- will drown in
the  growing one and, unless
$\mbox{Im } \sigma_n$ is relatively small, one cannot extract the
asymptotic amplitudes $x_{\rm out}$ and $x_{\rm in}$ from  the numerical data.  

Several methods have been devised to deal with this problem (Kokkotas and
Schutz 1992; Leins et al 1993; Andersson et al 1995). Most of these
were taken over from the study of black-hole quasinormal modes.
However, none of  these methods is suitable for 
coupled equations. For example, while the WKB method of Kokkotas
and Schutz (1992) can readily
be used for a single second order ODE it is very difficult
to use it for two coupled equations. In 
the slow-rotation problem one must deal with coupled equations. 
As long as the focus is on the spectral 
properties of the axial perturbations there is a way around this: By moving
the term in the right hand side of (\ref{ax1}) over to the left hand side and replacing
$X^{(0)}$ with $X^{(1)}$ we only introduce an error of order $\epsilon^2$. Thus, the
new equation --- that can be approached in the standard way --- is consistent
to $O(\epsilon)$. This trick was used by Kojima in his study of 
the polar problem (Kojima 1993b), but I have chosen not to use it here.

Instead, I have devised a ``new'' method to suppress the 
exponential divergence of the eigenfunctions. This method has the advantage that 
 it will remain useful
also for manifestly coupled equations. 
The integration of (\ref{ax0}) and (\ref{ax1}) is simply done using
complex-valued $r$ in the exterior spacetime. One would expect the 
exponential growth in 
$X\sim e^{\pm i \sigma r_\ast}\approx e^{\pm i \sigma r}$ to be 
suppressed if the integration 
is performed along a straight line such that $Arg(r)=-Arg(\sigma)$. 
Numerical tests show that this 
is indeed the case, and what remains is an (almost) purely oscillating 
function of $r$.
This function can be integrated with satisfactory precision from the 
surface of the star out to a large value of $|r|$. 
The numerical solution is then matched to an expansion of form
$e^{\pm i\sigma r_\ast} \sum a_n (1/r)^n$ at the final 
point of integration. This yields the values of the asymptotic
amplitudes  $x_{\rm out}$ and $x_{\rm in}$ for any given 
complex frequency.
Given these amplitudes one can iterate to
find a frequency such that $x_{\rm in}(\sigma_n)=0$, and identify a
quasinormal mode of the star. 

To test the reliability of the new integration method
I repeated the polar-mode calculations of Andersson, Kokkotas and Schutz (1995).
Most of their mode-frequencies for a non-rotating polytrope
could be reproduced with satisfactory precision. However, the new scheme
is not able to get to the few $w$-modes that are extremely rapidly damped.
Moreover, the new integration
approach is not accurate enough to enable iteration
for very slowly damped modes, like the $f$-mode, 
in contrast to the method used by
Andersson, Kokkotas and Schutz (1995). 
This is not a serious drawback, however. Slowly
damped modes can be identified by a simple
technique based on results for real frequencies [
the details of this method, which is an 
alternative to the standard resonance method (Chandrasekhar and Ferrari 1991b),
will be described
elsewhere].

\section{The effect of rotation on the axial modes}

In the non-rotating case the symmetries of the 
perturbation equations dictate that the stellar pulsation modes be 
symmetric
with respect to the imaginary $\sigma$ axis. That is, if $\sigma_n$ 
corresponds to
a mode then another mode is associated with $-\bar{\sigma}_n$
(the bar denotes complex conjugation). When the star
is rotating this symmetry is broken. Each of the modes for a non-rotating star
splits into $2l+1$ distinct modes, each corresponding to a different value of 
$m$ (cf. the Zeeman
splitting in quantum problems), and the symmetry changes in
such a way that  $\sigma_{nm}$ is now associated with $-\bar{\sigma}_{n-m}$.

The angular dependence of the perturbation functions is expressed 
in terms of
spherical harmonics (cf. Kojima 1992). Specifically, 
the dependence on the azimuthal angle is $e^{im\varphi}$. 
Since the time-dependence is assumed to be $e^{-i\sigma t}$ 
positive frequency ($\mbox{Re }\sigma>0$) modes for $m>0$ 
will be prograde relative to the 
stellar fluid and 
$m<0$ modes will be retrograde. Typically, 
retrograde modes will be destabilized by increased 
rotation, and the 
corresponding damping rates should decrease. 
For the prograde modes rotation will have a stabilizing effect, and
their damping rate should therefore increase. 
The reason for this is the central
mechanism of the Chandrasekhar-Friedman-Schutz instability (Chandrasekhar 1970;
Friedman and Schutz 1978): 
Suppose that a mode has a frequency such that it is retrograde
 according to an observer in the fluid, but because of 
the dragging of
inertial frames the mode appears prograde according to a 
distant observer. 
The distant observer
--- who measures energy according to his own time-line --- 
finds that the mode carries positive angular momentum away from the star.
But since the perturbed fluid actually rotates slower than it would in  
absence of the perturbation the angular momentum of the retrograde mode
is negative. The emission of gravitational waves makes
the angular momentum increasingly negative and leads to an instability.
 The same argument can be used
to explain the destabilization of retrograde modes in general.

A nonrotating perfect fluid star has no (non-trivial) axial pulsation modes
in Newtonian theory. All the familiar fluid pulsation modes (the $f$, $p$ 
and $g$ modes) are associated with polar perturbations. But in the 
relativistic picture an infinite
set of axial modes exist (Chandrasekhar and Ferrari 1991c; Kokkotas 1994). 
These are gravitational wave $w$-modes 
that exist because of the dynamic spacetime (see, for example, 
Kokkotas and Schutz 1992; Andersson et al 1996).
The fundamental $w$-mode of a non-rotating neutron star typically
has a pulsation frequency of the order of 10 kHz and damps out in a fraction of a 
millisecond. The modes become slower damped for
more relativistic models. Furthermore, it is possible that 
an increased rotation rate will lead to relatively slowly damped (or even
unstable)
$w$-modes since retrograde modes should become
slower damped as the rotation of the star increases. 

\begin{figure}[htb]
\epsfysize=7cm
\centerline{\epsffile {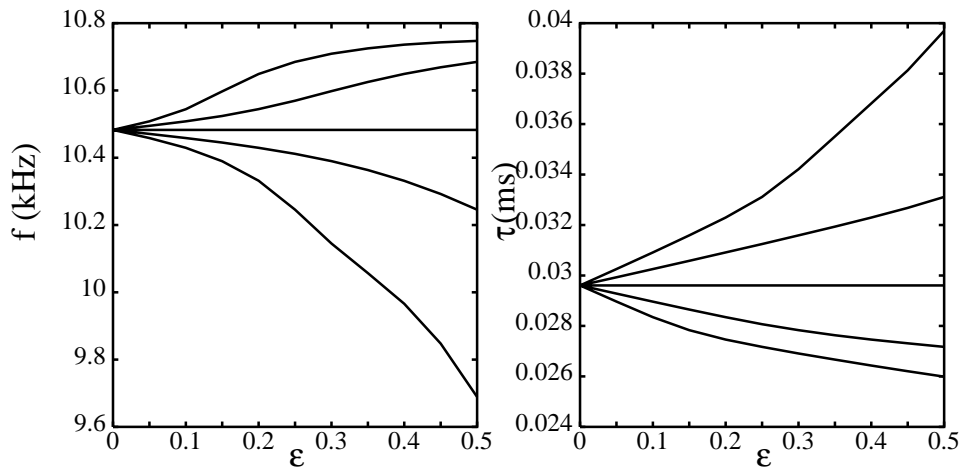}}
\caption{The effect of rotation on the slowest damped axial
gravitational wave mode for a  $\Gamma=2$ polytrope. This specific
star has central density $3\times10^{15} \mbox{g/cm}^3$, which leads to
$R=8.861$ km and $M=1.869 \mbox{ km }= 1.27 M_\odot$. 
The left panel shows the 
pulsation frequency of the star in kHz as a function of the 
dimensionless rotation parameter $\epsilon \equiv \Omega
\sqrt{R^3/M}$. For this stellar model $\epsilon \approx 0.2$ would
correspond to 
a pulsar with period 2 ms.  
The data corresponds to $l=2$ and (from bottom to top)
$m=-2,-1,0,1,2$. 
The right panel shows the e-folding time of the
mode in ms as function of $\epsilon$. Here the curves correspond to 
(from top to bottom) $m=-2,-1,0,1,2$. The $m<0$ modes are retrograde
while
the $m>0$ modes are prograde according to an observer in the
stellar fluid.}
\label{fig1}\end{figure}

I have performed a sample of calculations for the polytropic equation of state
\begin{equation}
p = 100 \mbox{km}\ \rho^2 \ .
\label{polyt}\end{equation}
Figure~\ref{fig1} shows results for the slowest damped
axial $w$-mode of a stellar model with central density 
$\rho_c = 3\times10^{15} \mbox{g}/\mbox{cm}^3$. This leads to
$R=8.861$ km and $M=1.869 \mbox{ km } = 1.27 M_\odot$, i.e. reasonable 
parameters for a neutron star.  The
results shown in Figure~\ref{fig1} nicely illustrate the effects of rotation
on the stellar pulsation modes. As anticipated, each mode for the 
non-rotating star splits into $2l+1$ distinct ones. The prograde modes
become shorter lived as the rotation rate increases, while the 
damping rate of the retrograde modes decreases.

As already mentioned there are no non-trivial axial pulsation modes
of a non-rotating star in Newtonian theory. But there is a set of
Newtonian axial modes that are degenerate at zero frequency. As soon
as the star rotates they become non-zero. These are the so-called
$r$-modes --- that are analogous to Rossby waves in the Earth's oceans ---
and Newtonian calculations to first order in rotation show that they
have frequencies of order $\Omega$ (Papaloizou and Pringle 1978;
Saio 1982). Newtonian calculations also show that the $r$-modes 
remain degenerate at first order in $\Omega$. To unveil the detailed
behaviour or the radial eigenfunctions and break the degeneracy one must proceed
to higher orders (Smeyers 1980; Provost et al. 1981; Smeyers et al. 1981;
Saio 1982). 

The $r$-modes have not previously  been
calculated in relativity, and the result of the present
calculation is perhaps surprising: The relativistic
analogue of the $r$-modes are unstable for \underline{all} rotating stars.
This result does, however, make perfect sense. A Newtonian calculation
to first order in $\Omega$ yields the $r$-mode frequency (Papaloizou and
Pringle 1978; Saio 1982);
\begin{equation}
\sigma_r =  m\Omega \left( 1 - {2 \over l(l+1) } \right) \ ,
\end{equation}
according to a distant observer. That is, these modes are retrograde
relative to the fluid but will appear prograde at infinity. 
Such modes should (as already discussed) be destabilized
by the emission of gravitational waves, and since the $r$-modes are
initially (for a non-rotating star) situated at $\sigma_r = 0$ they should become
unstable as soon as the star spins up. 

A sample of results for the $r$-modes and the polytropic equation of
state (\ref{polyt}) are shown in Table~\ref{table1}.
The tabulated data are for the same stellar model that was used
to generate the $w$-mode data in Figure~\ref{fig1}  (central density 
$\rho_c = 3\times10^{15} \mbox{g}/\mbox{cm}^3$). 
It is interesting to note that the growth-times for these modes
(represented by the ``e-unfolding'' time $\tau_e = 1/\mbox{Im } \sigma_n$) 
are rather short. 
For example, for a millisecond pulsar the amplitude of the
modes listed in Table~\ref{table1}
would multiply by a factor of three in a few seconds.

The $r$-mode growth rates given in Table~\ref{table1} should, however, 
be viewed with some caution. The reason for this is the simple fact that, 
for small frequencies $\sigma \sim \Omega$, the 
perturbation equations
(\ref{ax0}) and (\ref{ax1}) explicitly include terms that are formally
of second order in $\epsilon \sim \Omega$. As one can readily
deduce from the data in Table~\ref{table1} the numerical 
results scale with the
rotation parameter as $\mbox{Im } \sigma_r \sim \epsilon^2$ as  
$\epsilon\to 0$. Hence, the 
estimated growth rates for the $r$-modes are due to the included 
second order terms --- the eigenfrequency
is real to first order in $\epsilon$. That is, the presented results for the 
imaginary parts of the eigenfrequencies are not obtained through a 
consistent calculation. It is possible (likely) that the results will change 
considerably once the calculation is extended to order $\epsilon^2$.

The numbers listed in Table~\ref{table1} are nevertheless interesting.
They provide the first suggestion that the $r$-modes are unstable
in relativity. An argument based on the oscillation 
frequencies of the modes (which are of order $\epsilon$ and
should be obtained correctly in the calculation) and the mechanism
behind the Chandrasekhar-Friedman-Schutz instability clearly establishes
the existence of this instability. 
It is also far from clear how ``wrong'' the growth rates listed in
Table~\ref{table1} are. The error seems to be of the same
order as the result itself, but this does not necessarily 
mean that the obtained result
is useless 
as an order-of-magnitude estimate. What is clear at the moment is that
a calculation that is consistent to second order in $\epsilon$,
or an alternative estimate of the growth rate of the instability, is urgently
needed. It will either show that $\mbox{Im } \sigma_r \sim \epsilon^2$ or
establish that the true damping enters at higher orders. If the latter is the
case a slow-rotation calculation of these growth rates seems very
difficult. 

A second-order calculation would also be interesting since 
the degeneracy of the $r$-modes should be broken at order 
$\epsilon^2$.
Then $r$-modes with a different number of nodes in the radial eigenfunctions
should lead to distinct eigenfrequencies (Saio 1982).

\section{Discussion}

I have presented the first numerical results for 
axial pulsation modes of a slowly
rotating relativistic star. The calculation includes all terms of
first order in $\epsilon \equiv \Omega
\sqrt{R^3/M}$. It  includes
effects due to the coriolis force, but neglects effects due to the centrifugal 
flattening of the star. Such effects will appear at order $\epsilon^2$
and higher, and may well have a considerable effect on the various 
stellar pulsation modes.

It has been shown that increased
rotation has a stabilizing effect on prograde modes, while retrograde
modes are destabilized. This was expected from the mechanism behind
the familiar Chandrasekhar-Friedman-Schutz instability (Chandrasekhar 1970;
Friedman and Schutz 1978). Specifically, it was demonstrated 
that rotation affects the axial gravitational wave $w$-modes in the
anticipated way. 

I have also presented  the first
relativistic results for the so-called $r$-modes.
The somewhat surprising result that these modes are unstable 
due to 
the emission of gravitational radiation
even for the slowest rotating stars is interesting for many reasons.
First of all, axial modes have not been considered in previous studies
of stellar mode-instabilities (see, for example, Friedman 1983;
Lindblom 1986; Stergioulas and Friedman 1997). 
The present result shows that they 
are relevant, and that more detailed work is required in order for us
to determine whether the unstable $r$-modes are 
astrophysically important. 

We need to establish if this new class of instabilities
is stronger or weaker than the familiar instability associated with (say) the
$f$-mode of a rapidly rotating star. The answer 
depends on the relative growth times of the various unstable modes. As already 
mentioned, the numerical results presented here (cf. Table~\ref{table1}) 
should be used with considerable caution, but they will have to
serve as a guideline
until more reliable (higher order) data become available. 

If the unstable $r$-modes are to be of astrophysical interest they
must grow on a time-scale much shorter than that associated
with the damping due to viscosity, cf.
the instability associated with the $f$-mode that is
estimated to limit the rotation period of a $10^7$ K neutron star 
at the level of 1-2 ms (Cutler and Lindblom 1987).
To make a similar estimate based on the instability
growth rates obtained in this paper
is probably meaningless, especially since these results are expected
to change
(probably by several orders of magnitude) once more detailed 
studies are done. 

The $r$-mode instability is conceptually interesting, especially since it
 is active for \underline{all}, even the slowest, rotating stars. 
It is also possible that this new class of unstable modes of 
rotating relativistic stars is of considerable astrophysical relevance. The 
$r$-mode instability could, for example, limit the rotation
period of nascent neutron stars. Whether this is the case or not is 
obviously unclear at present. It is a challenge for future work to provide 
an answer to such questions.

\acknowledgments

It is a great pleasure to acknowledge enlightening discussions with (and/or 
helpful comments from) 
John Friedman, Yasufumi Kojima, Kostas Kokkotas, Norman Lebovitz, Lee Lindblom,
and Bernard Schutz. I am also grateful to the Albert Einstein
Institute in Potsdam for generous hospitality.

\pagebreak

\begin{table}[h]
\caption{r-modes for a $\Gamma=2$ polytrope. This specific
star has central density $3\times10^{15} \mbox{g}/\mbox{cm}^3$, which leads to
$R=8.861$ km and $M=1.869 \mbox{ km } = 1.27 M_\odot$. 
 All these modes are unstable due to the emission of gravitational waves. 
 The dimensionless
rotation parameter $\epsilon$, the rotation period $2\pi/\Omega$, the
pulsation frequency $f$ and the ``e-unfolding'' time $\tau_e$ of the modes
are listed.
The data are for $l=2$ and $|m|=2$.}
\begin{tabular}{c|c|c|c}
 $\epsilon$  &  Period (ms) & $f$ (Hz) & $\tau_e$ (s) \\ \hline
   0.001 & 404 &       3.35 &  $1.36\times 10^5$ \\
   0.010 & 40.4  &       33.5  &   1388 \\
   0.025 & 16.2  &       83.7 &    248 \\
   0.05 &  8.08  &      167  &     87.0 \\
   0.10 &  4.04  &      336  &     34.6 \\
   0.15 &  2.69  &      505  &     11.4 \\
   0.20 &  2.02  &      675 &      3.24 \\
   0.25 &  1.62  &      848   &     0.990 \\
   0.30 &  1.35  &     1024   &     0.368\\
   0.35 &  1.15  &     1203   &     0.180 \\
   0.40 &  1.01  &     1387   &     0.105 \\
   0.45 &  0.898  &     1576   &     0.0625 \\
   0.50 &  0.808  &     1772   &     0.0367 \\

\end{tabular}
\label{table1}\end{table}

\pagebreak

\end{document}